\documentclass[aps,prd,10pt,twocolumn,nofootinbib,floatfix,
  noshowpacs,preprintnumbers]{revtex4-1}
\usepackage{bm}
\usepackage{epsfig}
\usepackage{amsmath}
\usepackage{tablefootnote}
\usepackage{color}
\usepackage[dvipsnames]{xcolor}
\usepackage[colorlinks=true,breaklinks=true]{hyperref}
\hypersetup{allcolors=[rgb]{0.0 0.0 0.6},linkcolor=[rgb]{0.75 0.05 0.05}}
\usepackage[normalem]{ulem}

\renewcommand\[{\left[}

\newcommand{\TRGB}{{\rm TRGB}}

\newcommand{\exclude}[1]{}

\begin{document}

\preprint{MPP-2020-106}

\title{Axion and neutrino bounds improved with new calibrations of the
  tip of the red-giant branch using geometric distance determinations}

\author{Francesco Capozzi}
\author{Georg~Raffelt}
\affiliation{Max-Planck-Institut f\"ur Physik
(Werner-Heisenberg-Institut), F\"ohringer Ring 6, 80805 M\"unchen, Germany}

\date{July 7, 2020, revised August 24, 2020}

\begin{abstract}
The brightness of the tip of the red-giant branch (TRGB) allows one to constrain novel energy losses that would lead to a larger core mass at helium ignition and thus to a brighter TRGB than expected by standard stellar models. The required absolute TRGB calibrations strongly improve with reliable geometric distances that have become available for the galaxy NGC~4258 that hosts a water megamaser and to the Large Magellanic Cloud based on 20~detached eclipsing binaries. Moreover, we revise a previous TRGB calibration in the globular cluster $\omega$~Centauri with a recent kinematical distance determination based on Gaia~DR2 data.  All of these calibrations have similar uncertainties and   they agree with each other and with recent dedicated stellar models. Using NGC~4258 as the cleanest extra-galactic case, we thus find an updated constraint on the axion-electron coupling of $g_{ae}<1.6\times10^{-13}$ and $\mu_\nu<1.5\times10^{-12}\mu_{\rm B}$ (95\% CL) on a possible neutrino dipole moment, whereas $\omega$~Centauri as the best galactic target provides instead $g_{ae}<1.3\times10^{-13}$ and $\mu_\nu<1.2\times10^{-12}\mu_{\rm B}$.  The reduced observational errors imply that stellar evolution theory and bolometric corrections begin to dominate the overall uncertainties.
\end{abstract}

\maketitle

\section{Introduction}

The evolution of a low-mass star as it ascends the red-giant branch (RGB) is driven by the growing mass and shrinking size of its degenerate core until helium ignites and the core quickly expands \cite{Kippenhahn:2012}. The abrupt transition to a much dimmer helium-burning star on the horizontal branch (HB) leaves a distinct discontinuity at the tip of the red-giant branch (TRGB). It has been used for several fundamental applications besides, of course, for testing stellar-evolution theory.

Our main interest is to use the TRGB as a particle-physics laboratory in the sense that the emission of new low-mass particles, notably axions or neutrinos with anomalous magnetic dipole moments, would provide additional cooling of the helium core, thus increase the core mass before helium ignites, and therefore lead to a brighter TRGB. Comparing the modified stellar models with empirical calibrations provides constraints on e.g.\ the axion-electron interaction strength $g_{ae}$ or the anomalous neutrino dipole moment~$\mu_\nu$ \cite{Dearborn:1985gp, Raffelt:1989xu, Raffelt:1990pj, Raffelt:1992pi, Catelan:1995ba, Castellani:1993hs, Viaux:2013hca, Viaux:2013lha, Straniero:2018fbv, Arceo-Diaz:2015pva, Diaz:2019kim}.

Traditionally these studies relied on globular-cluster stars and the main uncertainty derived from the adopted distances. So our reconsideration is motivated, in part, by recent kinematical distance determinations for several galactic globular clusters based on Gaia~DR2 data \cite{Baumgardt:2019}. This method is geometrical and does not rely, for example, on the HB brightness. The reported distances typically agree very well with the traditional ones~\cite{Harris:1996}, but have much smaller uncertainties.

A second motivation is the availability of new theoretical reference models explicitly for the purpose of TRGB calibration, including detailed error estimates~\cite{Serenelli:2017}. These authors find that for stellar parameters appropriate for the globular cluster M5, their calibration agrees perfectly with earlier models dedicated to M5 \cite{Viaux:2013hca} after one corrects for the treatment of screening of nuclear reaction rates relevant for conditions on the RGB.

\begin{figure}[b!]
  \includegraphics[width=0.88\columnwidth]{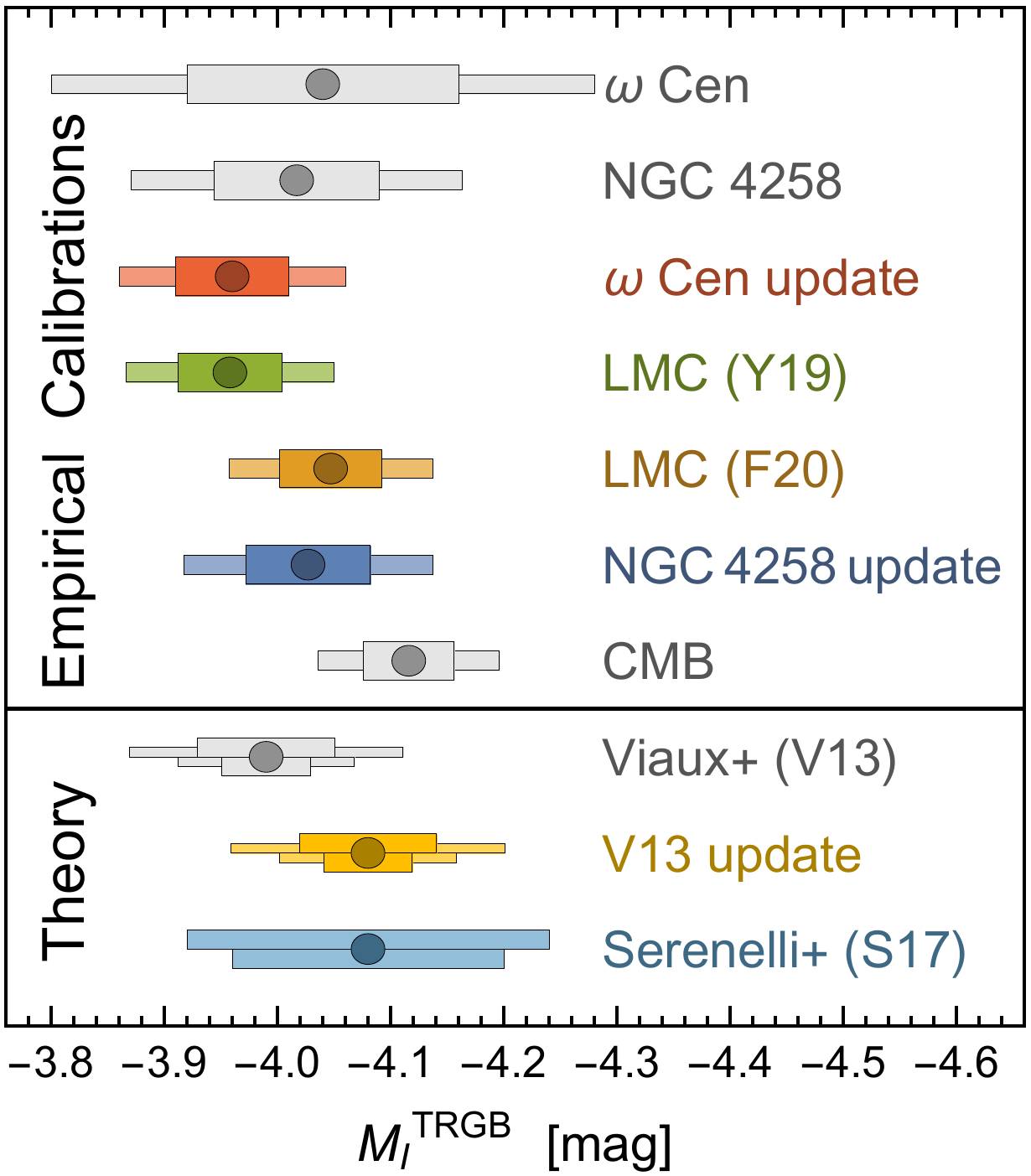}
  \caption{Summary of TRGB calibrations, showing 68\% and 95\% confidence intervals (see Table~\ref{tab:summary} for an annotated summary and Sec.~\ref{sec:TRGB-calibration} for a detailed discussion). The theoretical prediction of Serenelli et al.\  \cite{Serenelli:2017} instead uses a maximum uncertainty (see Sec.~\ref{sec:Theory}). The upper part of each theoretical error bar includes the contribution of the bolometric correction, whereas the
      lower part only includes uncertainties from stellar evolution theory. The cases marked ``update'' are our updates of previous results shown in gray. All gray-shaded cases are only shown for illustration and comparison.}\label{fig:summary}
\end{figure}

Another empirical TRGB calibration uses red giants in the haloes of galaxies as these also represent an old population of stars. In our own galaxy, eventually Gaia parallaxes should provide such a calibration, but current results are not yet competitive~\cite{Mould:2019}. However, the TRGBs of halo red giants in other galaxies have long been used as standard candles for distance determinations.\footnote{See e.g.\ the Introduction of Ref.~\cite{Jang:2017} for a historical review and their Table~7 for TRGB calibrations up to the year 2016.}
So another motivation for reconsidering the red-giant particle bounds is to use extragalactic TRGB calibrations for the first time in this context that can be seen as fundamental as determining the Hubble constant.

One still needs a distance anchor, notably the Large Magellanic Cloud (LMC) at a distance of around 50~kpc or the galaxy NGC~4258 (M106) at around 7.6~Mpc for which geometric distances are available. For the LMC, a recent precision distance estimate is based on 20~detached eclipsing binaries (DEBs) \cite{Pietrzynski:2019} that has been used for a TRGB calibration and determination of the Hubble constant \cite{Freedman:2019, Freedman:2020dne, Yuan:2019npk}.
While internal extinction within the LMC remains a systematic problem, this point is not an issue in NGC~4258. Moreover, it hosts a water megamaser that provides a geometric distance that has been updated very recently \cite{Reid:2019tiq}. We will use this significant improvement to update the NGC~4258-based TRGB calibration~\cite{Jang:2017}.

Each of these efforts provides the absolute $I$-band brightness $M_I^\TRGB$ of the TRGB at a reference color, here taken to be $(V-I)^\TRGB=1.8$~mag. The true variation of $M_I^\TRGB$ with $(V-I)^\TRGB$ or with metallicity is somewhat debated.\footnote{See e.g.\ Fig.~13 of Ref.~\cite{Jang:2017} for a compilation of previous findings and Ref.~\cite{Serenelli:2017} for a recent theoretical appraisal.} The empirical color dependence of  Ref.~\cite{Jang:2017} and the theoretical one of Ref.~\cite{Serenelli:2017} agree that the slope is very small in the relevant range around $(V-I)^\TRGB=1.8$ and the LMC calibrations of Refs.~\cite{Freedman:2019, Freedman:2020dne, Yuan:2019npk} assumed a vanishing slope as an input assumption. Indeed, the attraction of using the $I$-band TRGB brightness as a standard candle is precisely its weak, if any, dependence on color or metallicity. In this sense $M_I^\TRGB$ can be seen as {\em the\/} TRGB brightness, or it can be seen as the zero point at $(V-I)^\TRGB=1.8$ if a non-vanishing slope is considered.

Our first result, summarised in Fig.~\ref{fig:summary} and Table~\ref{tab:summary}
and discussed in Sec.~\ref{sec:TRGB-calibration}, is a compilation of those recent TRGB calibrations that are based on direct
geometric distances. Moreover, in Sec.~\ref{sec:Theory} we compare the theoretical reference models of Serenelli et al.\ \cite{Serenelli:2017}, which henceforth we will refer to as S17,\footnote{We will frequently refer to the following papers:
{\bf V13:}~Viaux et al.\ (2013) \cite{Viaux:2013hca}.
{\bf S17:}~Serenelli et al.\ (2017) \cite{Serenelli:2017}.
{\bf F19:}~Freedman et al.\ (2019) \cite{Freedman:2019}.
{\bf F20:}~Freedman et al.\ (2020) \cite{Freedman:2020dne}.
{\bf Y19:}~Yuan et al.\ (2019) \cite{Yuan:2019npk}.
}
with the earlier ones of Viaux et al.\ \cite{Viaux:2013hca},
henceforth V13, and the uncertainties identified by these
groups. These theoretical calibrations are also shown in
Fig.~\ref{fig:summary}, where in one case a Gaussian distribution of
errors is assumed, in the other a maximum range of uncertainty. The
stated zero points fortuitously are the same after the models of V13
have been corrected for the screening issue pointed out by S17.
Note that the errors on what we call theoretical predictions
  include a contribution from the empirical bolometric correction (BC),
  which is essential to compare with observational
  data. To distinguish this uncertainty from stellar evolution theory
  we show in Fig.~\ref{fig:summary} theoretical error bars
  with (upper) and without (lower) the error of the BC.

We continue in Sec.~\ref{sec:limits} with deriving limits on neutrino dipole moments and the axion-electron coupling by comparing the empirical calibrations of Sec.~\ref{sec:TRGB-calibration} with the modified theoretical models of V13 that were derived for the globular cluster M5, but equally apply to the other cases at the reference color $(V-I)^\TRGB=1.8$~mag. We finally wrap up in Sec.~\ref{sec:conclusion} with a discussion and summary.

\section{Empirical TRGB calibrations}

\label{sec:TRGB-calibration}

In this section, we turn to a panorama and assessment of those TRGB calibrations that are based on geometric distance determinations. We follow essentially a sequence of scale, beginning with the galaxy NGC 4258 at 7.6~Mpc all the way to globular clusters in our own galaxy.

\subsection{Galaxy NGC 4258 (M106)}

The galaxy NGC~4258 hosts a water megamaser that can be used as a geometric distance indicator. The latest estimate by Reid et al.\ (2019) \cite{Reid:2019tiq} based on 18~VLBI radio observation epochs and an improved analysis is \begin{equation}\label{eq:NGC4258-distance}
  d=7.576\pm0.112~{\rm Mpc}
  \quad\hbox{and}\quad
  \mu=29.397\pm0.032\,,
\end{equation}
where $\mu=5\log_{10}(d/{\rm pc})-5$ is the true distance modulus.
The most important change relative to their own previous value of $\mu=29.387\pm0.057$ \cite{Riess:2016jrr} is a reduction of the stated uncertainty by almost a factor of~2.

The most recent TRGB calibration using this galaxy was performed by
Jang and Lee (2017) \cite{Jang:2017} based on HST observations. They provide
their result in different filter systems. Removing the distance
modulus and its error from their final $VI$ result leads to
$I_0^\TRGB=25.364\pm0.045$ at their reference color $(V-I)_0^\TRGB=1.5$.
Their result includes foreground extinction, whereas
internal NGC~4258 extinction is neglected.  We shift the zero point to
our fiducial color of $(V-I)_0^\TRGB=1.8$ using the color dependence
provided in the last line of their Table~7. The implied dimming by 0.006~mag yields
\begin{equation}\label{eq:Jang-calibration}
  I_0^\TRGB=25.370\pm0.045.
\end{equation}
The itemized error budget is provided in the first column of
their Table~4 which includes $\pm0.023$~mag for the uncertainty of TRGB detection as an edge in the luminosity function. The modern version of the edge finding algorithm was proposed by Lee, Freedman and Madore (1993) \cite{Lee_1993}, who convolved the basic Sobel kernel with a binned luminosity function. Variations of this technique have then been applied by several authors. A historical review on the evolution of this technique is presented in Section~3.3 of Ref.~\cite{Freedman:2019}. The other main contributions to the error derive from photometric uncertainties and in particular the largest individual contribution of $\pm0.03$~mag from the F814W to $I$ filter transformation. We note that this stated error is much larger than the uncertainty of less than $\pm0.004$~mag
stated by Riess et al.\ (2016) \cite{Riess:2016jrr}.\footnote{For the transformation between
the $I$ band and the F814W filter, relevant for the wide-field camera (WFC3) of the HST,
Riess et al.\ (2016) \cite{Riess:2016jrr} in their Eq.~(11) provide
\begin{equation}\label{eq:filter-transformation}
{\rm F814W}=I-0.012-0.018\,\bigl[(V-I)-1.8\bigr]\,.
\end{equation}
The response of the two filters is shown e.g.\ in Fig.~2 of Y19.}

This issue only appears because we show all of our results, and compare with stellar evolution theory, in terms of the $I$-band. When comparing relative distances between galaxies based on HST observations this transformation would not appear. When comparing with stellar evolution theory, a large uncertainty derives from the bolometric correction (see Sec.~\ref{sec:Theory}). In a future dedicated analysis one could use directly the bolometric correction to F814W, avoiding a two-step transformation between observational and theoretical parameters.

An earlier TRGB calibration in NGC~4258 using HST
observations of a different field in this galaxy
was performed by Macri et al.\ (2006) \cite{Macri:2006} in a paper
otherwise devoted to Cepheid calibrations. Their edge-finding
algorithm turns up $I^{\rm TRGB}=25.42\pm0.02$ (see
their Fig.~20) with a statistical uncertainty very similar to the one
of Jang and Lee of $\pm0.023$.
Adopting an extinction of $A_I=0.025\pm0.003$ as
mentioned by Reid et al.\ (2019) \cite{Reid:2019tiq} yields
$I_0^{\rm TRGB}=25.395\pm0.02$ which agrees with
Eq.~\eqref{eq:Jang-calibration} on the $1\sigma$ level of the
statistical edge-finding error.

One may be tempted to combine these results to reduce the
statistical error. Reid et al.\ (2019) \cite{Reid:2019tiq} have
discussed these two determinations and have combined them on the F814W
level, but without reducing the error for fear of
correlations. Taking the average of these two
calibrations shifts the final zero-point of $M_I^\TRGB$ only by
0.01~mag.

In view of this small effect we prefer to avoid combining
heterogeneous results from different groups and rather stick to the
calibration worked out by Jang and Lee in great detail, including the
color variation, which itself causes shifts at the 0.01~mag level,
depending on the chosen reference color. So we adopt
\begin{equation}\label{eq:Jang-01}
  M_I^\TRGB=-4.027\pm 0.055
\end{equation}
by combining
Eq.~\eqref{eq:Jang-calibration} with the distance of
Eq.~\eqref{eq:NGC4258-distance}.

\subsection{Large Magellanic Cloud (LMC)}
\label{sec:LMC}

A much closer extragalactic target is the LMC for which a precise geometric distance determination has recently become available. Based on 20~detached eclipsing binaries (DEBs), Pietrzy{\'n}ski et al.\ (2019) found \cite{Pietrzynski:2019}
\begin{equation}\label{eq:LMC-distance}
  \mu=18.477\pm0.004_{\rm stat}\pm0.026_{\rm sys}=18.477\pm0.026,
\end{equation}
where the final error is dominated by systematics. The corresponding distance is $49.59\pm0.55$~kpc.

In analogy to NGC 4258, the TRGB is found with an algorithm to detect the corresponding edge in the luminosity function. A large data base is provided by OGLE-III \cite{Udalski:2008fa, Ulaczyk:2012cd}, a ground-based $VI$ survey of the LMC and the galactic bulge. This survey was originally motivated to search for gravitational lensing caused by astrophysical compact objects as dark matter candidates.

As our reference case, we follow the latest TRGB calibration of the Carnegie-Chicago group
(Freedman et al.\ 2019 and 2020 \cite{Freedman:2019,Freedman:2020dne}, henceforth F19 and F20).
They used
OGLE-III stars outside an ellipse surrounding the LMC bar and considered a color range
$(V-I)^\TRGB=1.6$--2.2 (see the white box in the upper-right panel of Fig.~5 in F20) which corresponds to the true color range 1.8--2.4 and they assumed that $M_I^\TRGB$ would not depend on $(V-I)^\TRGB$ across this range. Searching for an edge in the luminosity function, F19 found
\begin{equation}\label{eq:LMC-I-F19}
I^\TRGB=14.595\pm0.023.
\end{equation}
The data within the ellipse provides the same result, suggesting that crowding is not an important effect.

In an earlier study, Jang and Lee (2017) \cite{Jang:2017} used 10~fields of OGLE-III stars in the LMC and found $I^\TRGB_0=14.485\pm0.030$, which already includes their estimate of extinction $A_I$ of around 0.1~mag on average. Another determination based on OGLE-III data was performed by G\'orski et al.\ (2016) \cite{Gorski:2016} who used 17~fields and found $I^\TRGB=14.62\pm0.03$, compatible with
Eq.~\eqref{eq:LMC-I-F19} within stated uncertainties. More recently, Yuan et al.\ (2019) \cite{Yuan:2019npk} found $I^\TRGB_0=14.61$ without stating an uncertainty.
While all of these TRGB determinations used OGLE-III data in different and partly overlapping fields and are consistent with each other, we will not attempt to combine them and rather follow F19 and F20 as one specific reference case.

The main challenge is to determine the extinction $A_I$ that consists of a foreground part in the Milky Way and internal extinction within the LMC. Actually most of the work in F19 and F20 went into an updated estimate of extinction for which they found
\begin{equation}\label{eq:AI-F20}
  A_I=0.16\pm0.02\,,
\end{equation}
based on a comparison of TRGB colors between the LMC and the galaxy IC~1613 as well as the SMC.
Their corresponding TRGB calibration is
\begin{eqnarray}\label{eq:Freedman-01}
  M_I^\TRGB&=&-4.047\pm0.022_{\rm stat}\pm0.039_{\rm sys}
  \nonumber\\
  &=&-4.047\pm0.045
\end{eqnarray}
applicable at the color $(V-I)^{\rm TRGB}=1.8$~mag, which fortuitously agrees with the globular cluster M5 that we consider in Sec.~\ref{sec:M5} below.

However, Eq.~\eqref{eq:AI-F20} is the largest value for LMC extinction found in the literature and there is no consensus value at the present time. In an earlier paper of the Carnegie-Chicago group (Hoyt et al.\ 2018
\cite{Hoyt:2018ghz}), a reddening of $E(B-V)=0.03\pm0.03$ based on NIR colors of the TRGB was adopted, which is converted to $A_I=R_IE(B-V)=0.05\pm0.05$ using $R_I=1.76$ \cite{Bellazzini:2001}, although this very low value probably should be seen as being superseded by more recent estimates.

Another technique for estimating the extinction is based on the OGLE reddening maps \cite{Gorski:2020} using Red Clump stars as tracers and RR Lyrae stars in the LMC, leading to intermediate adopted values
\cite{Jang:2017, Yuan:2019npk, Nataf:2020}
\begin{equation}\label{eq:AI-Y19}
  A_I=0.10\pm0.02\,.
\end{equation}
An explicit recent TRGB calibration based on this estimate as well as the recent DEB distance is that of Y19 who found $M_{F814W}^\TRGB=-3.97\pm0.046$. Using the filter transformation of Eq.~\eqref{eq:filter-transformation} at the reference color $(V-I)=1.8$ we need to add 0.012~mag and adopt
\begin{equation}\label{eq:Y19}
  M_I^\TRGB=-3.958\pm0.046
\end{equation}
for this alternative calibration which is 0.09~mag dimmer than that of F20. The main difference arises from the different extinction value.

A thorough summary of recent reddening measurements using different stellar populations in the LMC is provided in Table 2 of Joshi and Panchal (2019) \cite{Joshi_2019}. In the same work they analysed the reddening distribution across the LMC using Cepheids provided by the OGLE IV survey. They find $E(B - V) = 0.091 \pm 0.050$ mag, which corresponds to $A_I=0.16$. Using the list of measurements provided by Joshi and Panchal, F20 have derived a probability distribution for the true value of the extinction in the LMC shown in their Fig.~2. It seems that the value indicated by Y19 lies  near the lower bound of the available measurements, whereas those preferred by F19, F20 and Joshi and Panchal are close to the center of the distribution.

The ongoing discussion about the LMC extinction is driven by the Hubble-tension debate. All else being equal, the Planck value of $H_0=67.4\pm0.5$ \cite{Aghanim:2018eyx} and using the F20 calibration of Eq.~\eqref{eq:Freedman-01} and Eq.~\eqref{eq:H0} would require $A_I=0.23\pm0.05$, even larger than Eq.~\eqref{eq:AI-F20}, whereas the Cepheid value of $H_0=74.03\pm1.42$ \cite{Riess:2019cxk} would nominally require $A_I=0.02\pm0.07$. Therefore, it depends on $A_I$ if the LMC-based TRGB calibration lends more support to one or the other $H_0$ value.

For deriving particle bounds we compare theoretical TRGB predictions with empirical calibrations. As seen in
Fig.~\ref{fig:summary}, F20 agrees rather well with theory, whereas Y19 is somewhat on the dim side. Additional energy losses can only brighten the TRGB, so using Y19 would lead to ``aggressive'' particle bounds. In this sense it is conservative for us to focus on the F20 calibration, of course keeping in mind that a more reliable $A_I$ determination could strengthen the particle bounds.

\subsection{Small Magellanic Cloud (SMC)}

F20 also considered the SMC using a DEB-based distance determination. After locating the edge in the luminosity function and applying an extinction correction, they found $M_I^\TRGB=-4.09\pm0.03_{\rm stat}\pm0.05_{\rm sys}=-4.09\pm0.06$. However, the adopted error may be too optimistic. The DEB distance is based on a two-step determination, combining the LMC distance with an earlier relative SMC-LMC measurement. More importantly, the uncertainty $\pm 0.007$~mag of their TRGB detection at $I^\TRGB=14.93$ would be much better than in the LMC.  On the other hand, G\'orski et al.\ (2016)
found $I^\TRGB=15.04\pm0.07$ where the large uncertainty owes to a large scatter between their five observational fields \cite{Gorski:2016}. Y19 reported $I^\TRGB=15.01$ without stating an uncertainty.
In view of these questions we will not consider the SMC as an independent TRGB calibration.

\subsection{Globular Cluster \boldmath{$\omega$} Centauri (NGC 5139)}

The most luminous globular cluster in our galaxy is $\omega$ Centauri. Its TRGB was calibrated by Bellazzini, Ferraro and Pancino (2001) \cite{Bellazzini:2001} and it was used to constrain novel energy losses of red giants by Arceo-D{\'\i}az et al.\ (2015) \cite{Arceo-Diaz:2015pva}. The $I$-band TRGB was found to be $I^{\rm TRGB}=9.84\pm0.04$~mag by searching for the corresponding edge in the luminosity function \cite{Bellazzini:2001}. This result must be corrected for the amount of extinction in the $I$ band for which Bellazzini et al.\ used $A_I=1.76\,(0.13\pm0.02)=(0.229\pm0.035)$~mag, so the true $I$-band brightness is $I_{0}^{\rm TRGB}=9.61\pm0.05$~mag.

A distance of $5.360\pm0.300$~kpc was determined with the detached eclipsing binary OGLEGC~17 by Thompson et al.\ (2001) \cite{Thompson:2001}. The corresponding distance modulus is $13.65\pm0.11$ and fixes the absolute brightness to be $M_I^\TRGB=-4.04\pm0.12$~mag  and thus reproduces the original calibration.\footnote{We use the linearized mapping between distance and distance modulus, as done in Bellanzini et al.\ \cite{Bellazzini:2001}. Adopting the logarithmic mapping would lead to a distance modulus of $13.656\pm0.121$.}

In order to update this previous calibration we first need to shift it to our fiducial color $(V-I)^\TRGB=1.8$ as the one from Bellazini et al.\ refers to $(V-I)^\TRGB=1.5$. S17's theoretical calibration shown in Eq.~\eqref{eq:Serenelli-01} suggests that the brightness difference between this and our fiducial color is $\delta M_I^\TRGB=0.01256$~mag, so the zero point would be dimmer by this amount. The empirical color dependence found by Jang and Lee (2017) \cite{Jang:2017} shown in the last line of their Table~7 and as a red line in their Fig.~13 suggests that this difference would be only 0.00609~mag. As a final shift to be applied to $I_0^{\rm TRGB}$ we choose the average of these two values, i.e., 0.0093 mag.

A second update is related to the extinction. Bellazzini et al.\ took $A_I$ from Ref.~\cite{Thompson:2001}, which in turn based their estimate on the dust map presented in Schlegel et al.\ (1998) \cite{Schlegel:1997yv}. Recently, Schlafly and Finkbeiner (2011) \cite{Schlafly_2011} have tested this map on the colors of stars with spectra in the Sloan Digital Sky Survey and they provided a table of conversion coefficients, which are meant to correct the predictions based on \cite{Schlegel:1997yv}. The conversion factor \cite{Schlafly_2011}\footnote{We take advantage of the online tables based on this publication provided at
\href{https://irsa.ipac.caltech.edu/applications/DUST/}{https://irsa.ipac.caltech.edu/applications/DUST/}} for the usual Landolt $I$ bandpass is $A_I/E(B-V)=1.505$, which multiplied by $E(B-V)=0.142$ from Schlegel et al.\ gives $A_I=0.214$. This value is 0.015~mag smaller than the one adopted in the original calibration of $\omega$~Centauri, so the zero point becomes
dimmer by this amount. Concerning the error we take 5\% of the extinction itself, as proposed by Schlafly and Finkbeiner (2011) \cite{Schlafly_2011}. Thus overall we adopt
\begin{eqnarray}\label{eq:M5-I}
  I_0^\TRGB&=&(9.84\pm0.04)_{\rm meas}-(0.214\pm0.011)_{\rm extinct}
  \nonumber\\
  &&{}+0.0093_{\rm color}
  \nonumber\\
  &=&9.635\pm0.041
\end{eqnarray}
for the apparent true $I$-band brightness at our reference color of $(V-I)_0^\TRGB=1.8$.

The most important modification comes from the distance. The canonical one in the catalog of Harris~\cite{Harris:1996}, based on the HB brightness, is 5.2~kpc without a specified uncertainty. It is perfectly consistent with the DEB distance cited earlier.

However, recently distances to selected galactic globular clusters were determined kinematically based on Gaia DR2 data by Baumgardt et al.\ (2019) \cite{Baumgardt:2019} (see their Table~3). For $\omega$ Centauri they found
\begin{equation}\label{eq:Cen-dist}
  d=5.24\pm0.05~{\rm kpc}
  \quad\hbox{or}\quad
  \mu=13.597\pm0.021,
\end{equation}
consistent with both distance determinations mentioned earlier, but with a much smaller uncertainty.

Taking into account the shifts due to color, extinction and distance, our updated zero point is
\begin{equation}\label{eq:Bellazzini-01}
  M_I^\TRGB=-3.96\pm0.05~{\rm mag},
\end{equation}
which is 0.08~mag dimmer than Bellazzini's orignal result, the shift coming mostly from the revised distance, but all of our corrections go in the same direction of making the TRGB dimmer.
Now, the distance is no longer the main source of uncertainty
for the TRGB calibration by $\omega$ Centauri.

\subsection{Globular Cluster M5 (NGC 5904)}

\label{sec:M5}

All TRGB calibrations discussed so far were based on finding a break in the luminosity functions of certain ensembles of stars, that however included populations with different chemical composition, age and mass, an issue that even applies to the globular cluster $\omega$ Centauri.  However, for our purpose of comparing the TRGB with theoretical models it may be cleaner to consider globular clusters where these parameters are more uniform, resulting, for example, in a clear separation of the red-giant branch (RGB) from the asymptotic giant branch (AGB). Of course, the disadvantage is the paucity of red giants near the TRGB, resulting in a significant uncertainty of the TRGB determination.

Motivated by these arguments, V13 studied the upper RGB in the globular cluster M5 in order to constrain novel energy losses in the degenerate cores before helium ignition. Their theoretical zero-point prediction is discussed in Sec.~\ref{sec:Viaux-Theory} below. Their philosophy was to identify the brightest RG and determine the statistical offset to the true TRGB. For the three brightest red giants they reported $I_{1,2,3}=10.329$, 10.363 and 10.420, with a typical uncertainty of $\pm0.023$, where they accounted for various sources of observational errors (photometry, crowding, etc). The color of the brightest star is approximately $(V-I)_1\approx 1.8$~mag.

If the core mass at helium ignition is increased by novel energy losses, the HB also brightens because zero-age HB stars will have larger core masses. Therefore, V13 avoided distances based on the HB brightness such as the canonical distance of 7.5~kpc in the catalog of Harris~\cite{Harris:1996}. Instead, they used the distance modulus $14.45\pm0.11$~mag ($7.76\pm0.39$~kpc) of Layden et al.\ (2005) \cite{Layden:2005} based on main-sequence fitting. We now use the kinematical distance of Baumgardt et al.\ \cite{Baumgardt:2019} of $7.58\pm0.14$~kpc, corresponding to $\mu=14.398\pm0.040$, similar to the Harris distance and also consistent with that from main-sequence fitting, but with a much smaller uncertainty.

Using a geometric distance, we also need to include extinction explicitly. Based on the update of Schlafly and Finkbeiner (2011) \cite{Schlafly_2011} we use an extinction of
$A_I=0.054\pm0.02$, where we have adopted a somewhat arbitrary estimated uncertainty.
So the brightest RGB in M5 is now found to have an absolute $I$-band brightness of
$M_I^{\rm 1st}=-4.123\pm0.050$. The dominant error still derives from the distance.

To estimate the difference between $M_I^{\rm 1st}$ and the true TRGB, V13 used the observed RGB population to estimate the evolutionary speed along the upper RGB and also included a brief hesitation at the TRGB suggested by the theoretical speed of evolution. With a Monte Carlo simulation they generated random realisations of the RGB with the same underlying distribution and in this way found the statistical distribution for the brightness difference $\Delta_{\rm tip}\geq0$ between the brightest star and the true TRGB. In this way they found $\langle\Delta_{\rm tip}\rangle=0.048$~mag and an rms variation of $0.058$~mag. While this extrapolation to the TRGB provides an asymmetric distribution, for simplicity we treat this error as a Gaussian uncertainty and finally find an updated calibration of $M_I^\TRGB=-4.17\pm0.08$, significantly brighter than the calibration from $\omega$~Centauri.

One obvious concern is if the brightest star in the list of V13 is indeed on the RGB and not an AGB contamination. In this context we notice that stars on the upper RGB/AGB tend to be long-period variables (LPV) with periods of 30--100 days \cite{Wood_1999,Wood_2000,Kiss_2006,Fraser:2008yp,Osborn_2017,Deras_2019}. For M5, this question was studied explicitly by Wehrung and Layden (2013) \cite{Wehrung:2013}. So the measured average brightness and color depends on the exact epochs of observation. The data used in V13 were based on the collected photometry by Stetson et al.\ available at that time. Meanwhile, an updated catalogue by Stetson et al.\ (2019) of homogeneous ground-based globular cluster observations has become available \cite{Stetson:2019}.\footnote{We use the 2020 update at \href{https://www.canfar.net/storage/list/STETSON/homogeneous/Latest_photometry_for_targets_with_at_least_BVI/}{https://www.canfar.net/storage/list \newline
/STETSON/homogeneous/Latest\_photometry\_for\_targets\_with\_
\newline
at\_least\_BVI/}}
The three brightest stars of V13 are now reported with
$I_{1,2,3}=10.295$, 10.344 and 10.436, which arise from an average of 118, 122 and 149 observations, respectively. The $(V-I)$ colors have also significantly changed.

The brightest star, in particular, identified by its equatorial coordinates (J2000) of
${\rm RA}=15^{\rm h}18^{\rm m}36.05^{\rm s}$ and ${\rm DE}=02^\circ
06'37.4''$ appears in the list of variable stars in M5 of Arellano
Ferro et al.\ (2016) \cite{Arellano:2015, Arellano:2016} as a
semi-regular late-type variable (SRA) named V50 with a period of
107.6~days. The reported average brightnesses
are $\langle V\rangle =12.15$~mag with a typical amplitude during a
cycle of $\pm0.37$~mag and $\langle I\rangle =10.27$~mag and a variation
amplitude of $\pm0.14$~mag, leading to a color index of $\langle V-I\rangle=1.88$ with $\pm0.23$ variations.

While the $I$ and $V$ amplitudes of the other stars near the TRGB are much smaller, an RGB/AGB discrimination would have to be reconsidered and a corresponding probability distribution developed. Clearly the uncertainty of the TRGB calibration would become larger than that based on our update of the V13 results. These issues have become important because the geometric distance uncertainty is now so small that such sources of uncertainty require much more careful attention. Therefore, we no longer use M5 as an independent TRGB calibration.
This situation suggests that one should consider more cases of individual globular clusters to reduce the impact of low-number statistics and the question of a star-by-star RGB/AGB separation near the TRGB.

\begin{table*}
\caption{Summary of empirical TRGB calibrations and implied bounds on
  the axion-electron coupling $g_{13}$.\label{tab:summary}}
\begin{tabular*}{\textwidth}{@{\extracolsep{\fill}}lllllllll}
  \hline\hline
  Target&Distance&$\mu$ \footnotemark[1]&$I^\TRGB$ \footnotemark[2]&$A_I$ \footnotemark[3]&$M_I^\TRGB$ \footnotemark[4]&Reference&\multicolumn{2}{l}{Axion Bound}\\
        &Method  &[mag]                 &[mag]                     &[mag]                 &[mag]                       &         &\quad 68\%&95\%\\
  \hline
  NGC 4258&Megamaser\footnotemark[5]&$29.397\pm0.032$&$25.395\pm0.045$&$0.025\pm0.003$ &$-4.027\pm0.055$&Update of \cite{Jang:2017}&\quad 0.79 &1.58 \\
  LMC&DEBs\footnotemark[6]&$18.477\pm0.039$&$14.595\pm0.023$&$0.16\pm0.02$ \footnotemark[7]&$-4.047\pm0.045$&F20 \cite{Freedman:2020dne}&\quad 0.81&1.58\\
  $\cdots$& $\cdots$               & $\cdots$           &  ---            &$0.10\pm0.02$ \footnotemark[7]&$-3.958\pm0.046$ \footnotemark[8]&Y19 \cite{Yuan:2019npk}&
  \quad0.62&1.25\\
  $\omega$
  Centauri&Kinematical\footnotemark[9]&$13.597\pm0.021$&$9.85\pm0.04$&$0.214\pm0.035$&$-3.96\pm0.05$&Update
  of \cite{Bellazzini:2001}&\quad 0.64&1.29\\
  \hline
\end{tabular*}
\footnotetext[1]{True distance modulus $\mu=5\log_{10}({\rm distance}/{\rm pc})-5$}
\footnotetext[2]{$I$-band brightness of the TRGB shifted to our
  reference color of $(V-I)^\TRGB_0=1.8$~mag}
\footnotetext[3]{Extinction}
\footnotetext[4]{Absolute $I$-band TRGB brightness:
  $M_I^\TRGB=I^\TRGB-A_I-\mu$}
\footnotetext[5]{Water megamaser with 18 VLBI radio observation epochs \cite{Reid:2019tiq}}
\footnotetext[6]{Detached eclipsing binaries \cite{Pietrzynski:2019}}
\footnotetext[7]{There is no consensus in the literature about the
  average extinction (foreground and internal) of the LMC, see Sec.~\ref{sec:LMC}. The two shown cases are taken to represent the plausible range.}
\footnotetext[8]{Y19 state $-3.97\pm0.046$ in the F814W filter.}
\footnotetext[9]{Kinematical distances of galactic globular clusters
  based on Gaia DR2 data \cite{Baumgardt:2019}}
\end{table*}

\subsection{Compound Globular Cluster}

In principle, this exercise was performed by F20 as an overall consistency check of their LMC calibration. They considered a selection of 11 galactic globular clusters and produced compound CMDs in the $JHK$ bands using 2MASS data. The relative distances were linked to the average HBs and/or RR Lyrae stars. The absolute distance was anchored to 47~Tuc based on DEB distances. While the result agrees with the other calibrations, it is the most uncertain of their cases and also more uncertain than the ones based on $\omega$~Centauri and M5, so we will not use it. Moreover, it involves a distance ladder to the observed stars, not a direct geometric determination.

In principle, one could use the list of globular clusters with good kinematical distances of Baumgardt et al.\ (2019) and repeat the exercise of Freedman et al.\ (2020). Also, recently a list of 50 globular clusters was used to constrain axions and neutrino dipole moments \cite{Diaz:2019kim}, however again relying on HB brightness distances.

Likewise, the bolometric TRGB brightness determinations in many globular clusters from the near infrared photometry of Ferraro et al.\ (1999) \cite{Ferraro:1999xa} and Valenti et al.\
\cite{Valenti:2004hm, Valenti:2004nf, Valenti:2007, Valenti:2010} are very interesting, but they did not provide an explicit $I$-band TRGB calibration and the uncertainties, especially those from the distances, are not clearly laid out. We find it too difficult to post-process these results for our present purpose where the uncertainty of $M_I^\TRGB$ is crucial, not only its zero point. The photo\-metry itself does not seem to be the limiting issue, but the distances are, so a clear discussion of the different sources of uncertainty for the final result is mandatory to be able to compete with the calibrations shown in Fig.~\ref{fig:summary}.

\subsection{Hubble Constant}

One main motivation for the TRGB calibration is to establish one rung in the cosmic distance ladder to determine the Hubble constant. On the basis of their LMC calibration, shown as our Eq.~\eqref{eq:Freedman-01}, F20 found $H_0=69.6\pm0.8_{\rm stat}\pm1.7_{\rm sys}=69.6\pm1.9~{\rm km}~{\rm s}^{-1}~{\rm Mpc}^{-1}$. The TRGB calibration was used to calibrate a sample of SNe~Ia which then take us to cosmological distances. We may compare this result with the cosmological determination of $H_0=67.4\pm0.5$ found by the Planck Collaboration \cite{Aghanim:2018eyx}. The two values agree at the $1.1\sigma$ level, meaning that they agree, but the CMB determination has a much smaller uncertainty.

The Cepheid-based value of $H_0=74.03\pm1.42$ \cite{Riess:2019cxk} is around $1.9\sigma$ larger than the TRGB-based one but still compatible. However, the Planck and the Cepheid values differ by around $4.4\sigma$, a discrepancy debated in the literature as the Hubble tension between local and large-scale $H_0$ calibrations. We will not pursue this topic here and simply note that there is no tangible Hubble tension between the Planck and TRGB results.

For illustration we can therefore turn these results around and ask: beginning with $H_0$ found by Planck, which TRGB calibration would we get after climbing down the cosmic distance ladder? The direct and inverted
LMC-based zero points of F20 are
\begin{subequations}\label{eq:H0}
\begin{eqnarray}
   \kern-2em H_0&=&69.6\pm1.24 + 32.05\,\left(4.047 +  M_I^\TRGB\right)\,,
   \\[1ex]
   \kern-2em M_I^\TRGB&=&-4.047\pm0.039+0.0312\,(H_0-69.6)\,,
\end{eqnarray}
\end{subequations}
where here and henceforth $H_0$ is understood in units of ${\rm km}~{\rm s}^{-1}~{\rm Mpc}^{-1}$. These expressions are the linearised versions of mapping between distance and distance modulus and we note that $H_0$ plays the dimensional role of an inverse distance.  A larger $H_0$ implies a smaller distance to a galaxy of fixed redshift and thus a dimmer TRGB.

In order to explain our stated uncertainties we note that $M_I^\TRGB\leftrightarrow H_0$ through SNe~Ia accrues an error even if the input information were exact. Starting from Eq.~\eqref{eq:Freedman-01} F20 find $H_0=69.6\pm1.9$, but if we propagate only the error of $0.045$~mag through Eq.~\eqref{eq:H0} without SNe~Ia error we find only $\pm1.44$, so the difference $\sqrt{1.9^2-1.44^2}=1.24$ must come from the transition through SNe~Ia. Inverting this result means that an uncertainty of 1.24 in $H_0$ translates into one of $0.0386$~mag in $M_I^\TRGB$. The cosmological determination by the Planck Collaboration \cite{Aghanim:2018eyx} is $H_0=67.4\pm0.5$, implying
\begin{equation}\label{eq:Hubble}
  M_I^\TRGB=-4.116\pm0.040\,.
\end{equation}
Of course, this calibration depends on the assumption that the local $H_0$ is identical with the large-scale value. Therefore, we will not use this result to derive particle bounds and only show it for illustration.

\subsection{Halo of the Milky Way}

In principle, the TRGB can be calibrated with the halo red giants in
our own galaxy, although this approach has only become thinkable with
the precision parallaxes provided by the Gaia astrometric
satellite. Based on the Data Release 2, a first attempt was made by
Mould et al.\ (2019) who show in their Fig.~1 a CMD of their stars
that were chosen in the direction of the South Galactic
Pole. Unfortunately, the parallax errors shown in their Table~1 for
the brightest stars translate to distance uncertainties on the order
of 10--30\%, also shown as errors on the implied $M_I$ values in their
Fig.~1. The distances are distributed in the range 4--14~kpc with a strong peak around 6~kpc.
Because of the large individual distance errors, these authors did not succeed in deriving a quotable TRGB calibration and instead have used their work as a consistency check with previous TRGB calibrations and as a proof of principle that the future Gaia DR3\emph{} will allow for a quantitatively competitive TRGB calibration. At this time we cannot use this approach to derive particle bounds or a value for the Hubble constant.

\subsection{Summary of Calibrations}

We next summarise these results in Fig.~\ref{fig:summary}, showing for each case the 68\% and 95\% confidence intervals, and summarize more details in the annotated Table~\ref{tab:summary}. For comparison we also show the $\omega$~Centauri and NGC~4258 calibrations that were based on earlier distances and for illustration our CMB calibration that we backward-engineered from F20's $H_0$ determination.

We also anticipate our update of the theoretical calibration by V13 that will be given in Eq.~\eqref{eq:Viaux-01}, whereas the one by S17 in Eq.~\eqref{eq:Serenelli-02}. The main difference between these cases is the treatment of error, where S17 gave a maximum error.

All of the empirical results agree well with each other within the stated uncertainties. It is actually reassuring that the scatter between different cases is roughly commensurate with the stated uncertainties which thus look realistic, i.e., the agreement is not ``too good to be true.'' On the other hand, the stated nominal $1\sigma$ uncertainties are always very nearly $\pm0.05$ despite very different sources of errors, so this agreement on overall error is somewhat striking.

In any event, we do not think that combining these calibrations to achieve a better estimate for the true value with reduced error would be meaningful.
Some of the variations are outright systematic, notably the two LMC calibrations which use discrepant values of extinction. On paper, the cleanest cases are NGC~4852 and $\omega$~Centauri, but even taking their average would probably obfuscate the current situation rather than adding meaningful information concerning the $H_0$ tension or particle emission from red-giant cores.

\section{Theoretical TRGB calibration}

\label{sec:Theory}

\subsection{Serenelli et al.\ (2017) -- S17}

S17 \cite{Serenelli:2017} have recently provided a new theoretical TRGB calibration, with a focus on various uncertainties that could affect the result. In particular from the comparison of two independent stellar evolution codes they derive a set of physical and numerical ingredients, which both minimize the results of the two codes and provide state-of-the-art inputs appropriate for low-mass stellar models. They recommend such models as a reference for further applications of the TRGB in astrophysics. In the color range $1.40<(V-I)^\TRGB<2.40$ they find
\begin{equation}\label{eq:Serenelli-01}
   M_I^\TRGB=-4.090 + 0.017\,{\rm Col}+0.036\,{\rm Col}^2\,,
\end{equation}
where ${\rm Col}=\bigl[(V-I)^\TRGB-1.4\bigr]$. At $(V-I)^\TRGB=1.8$, relevant for the empirical M5 and LMC calibrations, the zero-point is $M_I^\TRGB=-4.077~{\rm mag}$. Here the slope is $dM_I^\TRGB/d(V-I)^\TRGB=0.046$, so the color dependence is very small.

Concerning uncertainties, these authors discuss in their Sec.~5.1 a list of input parameters that are varied between extreme assumptions. Taking them to add coherently in one and the other direction, they find a full width of the predicted  $M_I^\TRGB$ range of 0.25--0.30~mag. In their Sec.~6 they consider specifically the color relevant for the globular cluster M5 and recommend an error of $\pm0.12$~mag, which we interpret as a maximum range.

However, this does not include the uncertainty of the bolometric correction (BC) for which they use MARCS \cite{Gustafsson:2008} which is compared to other BCs in their Fig.~8. At $(V-I)^\TRGB=1.8$ the spread between different BCs is around 0.1~mag. S17 explicitly find that for M5 the predicted $M_I^\TRGB$ becomes brighter by 0.07~mag if one uses the Worthey and Lee \cite{Worthey:2011} BC instead of MARCS. In the spirit of adding theoretical uncertainties coherently we add 0.04~mag to their recommended half-width error of $\pm0.12$. Therefore, we interpret S17's prediction for the zero-point calibration at $(V-I)^\TRGB=1.8$ as
\begin{equation}\label{eq:Serenelli-02}
   M_I^\TRGB=-4.08\pm\left(0.12_{\rm models}+0.04_{\rm BC}\right)_{\rm max}\,,
\end{equation}
where we interpret the uncertainty as a maximum range.

\subsection{Viaux et al.\ (2013) -- V13}

\label{sec:Viaux-Theory}

A similar study was performed earlier by V13 \cite{Viaux:2013hca} specifically for the globular cluster M5. For their fiducial case, they give in their Eq.~(6) $M_I^\TRGB=-4.03$~mag, using the Worthey and Lee BCs. S17 find a value of $-4.14$~mag using the same BC, different from the benchmark MARCS assumed in the previous section (see also the discussion in their Sec.~6). These authors argue that V13 should have used in their numerical calculations the intermediate screening regime in nuclear reaction rates instead of the Salpeter formulation of weak screening. This modification would make the V13 models brighter by 0.09~mag, so their zero point should be $M_I^\TRGB=-4.12$~mag, very close to S17's result. Thus for common input physics one finds a robust prediction despite many small differences in detail and despite using different codes.

Concerning the uncertainty of the prediction, V13 also listed a large number of possible input variations in their Table 4 as well as the assumed BC uncertainty given in their Eq. (9). Some of these ranges produce an asymmetric effect, notably the last two lines in Table 4 (equations of state and mass loss). This asymmetry shifts the prediction by $0.039$~mag in the dim direction. Apart from the screening prescription, this shift is the missing bit of difference to the models of S17 that was left unexplained in their Sec.~6. So after correcting for the treatment of screening, apparently there are no unexplained differences between the predictions.

Concerning the formal error, V13 proposed to combine the systematic errors in quadrature and to assume a top-hat distribution for each individual one, leading to an rms error of $\pm0.039$ from the uncertainties in Table~4. The BC error of $\pm0.08$, assumed to represent a maximum range with a top-hat distribution, corresponds to an rms uncertainty of $\pm0.08/\sqrt{3}=\pm0.046$. Combining these errors in quadrature and after applying the screening correction, V13's updated calibration is
\begin{equation}\label{eq:Viaux-01}
   M_I^\TRGB=-4.08\pm0.06_{\rm rms}\,.
\end{equation}
Although the two groups made some different choices in detail, their
zero points fortuitously coincide.\footnote{Notice in particular that V13
  used the BC of  Worthey and Lee whereas S17 used MARCS, so the exact
  coincidence of the two $M_I^\TRGB$ zero points is indeed coincidental.}

This calibration agrees very well with the empirical results. Therefore, in order to derive limits on axions or to provide a rung in the cosmic distance ladder, the main question is how to deal with systematic uncertainties, where the two groups have adopted different philosophies. However, we note that a top-hat distribution of half-width $s$ has an rms width of $s/\sqrt{3}$, so a maximum error of $\pm0.16$ corresponds to an rms error of $\pm0.09$. In other words, the formal uncertainties of the two calibrations are not as different as it may seem. The maximum error of S17 nominally corresponds  to a $2.7\sigma$ error of V13.

\subsection{Comparing with Empirical Calibrations}

Figure~\ref{fig:summary} suggests that the empirical calibrations and
theoretical predictions agree very well within the stated
uncertainties. To quantify this comparison we
ask for the allowed range of a possible mismatch
$\Delta M_I^\TRGB=M_I^{\rm Empirical}-M_I^{\rm Theory}$ between a
given calibration and theoretical prediction.

The uncertainty of this offset depends on how to treat the
uncertainties of the theoretical prediction. V13 have argued that one
should combine many sources of systematic errors in quadrature and
interpret the final result as a Gaussian error. We have already
combined statistical and systematic uncertainties of the empirical
result in quadrature and have interpreted the combined error as a
Gaussian uncertainty. Combining the errors of the empirical
calibrations and of the theoretical prediction in quadrature, the
offset (in mag) is constrained by
\begin{equation}\label{eq:Viaux-offset}
  \Delta M_I^\TRGB=
  \begin{cases}
    +0.05\pm0.08 &\quad\hbox{NGC 4258}\\
    +0.03\pm0.08 &\quad\hbox{LMC (F20)}\\
    +0.12\pm0.08 &\quad\hbox{LMC (Y19)}\\
    +0.12\pm0.08 &\quad\hbox{$\omega$ Centauri}
  \end{cases}
\end{equation}
The errors are dominated by that of the prediction. All calibrations are slightly dimmer than expected, but perfectly consistent within uncertainties.

If instead one combines the stellar-evolution errors to provide a
possible maximum range in the spirit of S17, it is not completely
clear on how to combine the empirical and theoretical errors. We
simply note that the maximum range of $\pm0.16$~mag is much larger
than the calibration errors of around $\pm0.05$~mag, so one could
essentially neglect the latter. In this case the maximum allowed
mismatch between theory and calibration of $\pm0.16$~mag corresponds
roughly to a $2\sigma$ range of
Eq.~\eqref{eq:Viaux-offset}. Therefore, the practical difference of
the two philosophies becomes small because any substantial conclusion
based on Eq.~\eqref{eq:Viaux-offset} would be based on a $2\sigma$
interval at least, so the nominal uncertainties are not very
different.

\section{Particle Bounds}

\label{sec:limits}

\subsection{Testing Standard Neutrino Emission}

In the context of the Standard Model neutrino emission from the
degenerate helium core is dominated by plasmon decay,
$\gamma_{\rm pl}\to\nu\bar\nu$, a process that can not be tested in
the laboratory. The calculated emissivity is thought to be precise on
the 5\% level. In their Sec.~5.11, V13 showed that changing standard
neutrino emission by $\pm5\%$ changes $M_I^\TRGB$ by
$\mp0.013$~mag. If we denote with $F_\nu$ a fudge factor multiplying
standard neutrino emission, the constraints of Eq.~\eqref{eq:Viaux-offset}
can be interpreted as
\begin{equation}\label{eq:fudgefactor}
  F_\nu=
  \begin{cases}
    0.80\pm0.31 &\quad\hbox{NGC 4258}\\
    0.87\pm0.29 &\quad\hbox{LMC (F20)}\\
    0.53\pm0.29 &\quad\hbox{LMC (Y19)}\\
    0.54\pm0.30 &\quad\hbox{$\omega$ Centauri}
  \end{cases}
\end{equation}
In other words, standard neutrino emission is confirmed with
the expected rate, but the test is not very precise.

\subsection{Predicted Brightness Increase by Neutrino Dipole Moments}

If the core of a red giant before helium ignition suffers energy
losses in addition to the usual neutrino emission, a larger core mass
is required so that also the brightness $M_I^\TRGB$ increases. Viaux
et al.\ (2013) \cite{Viaux:2013hca, Viaux:2013lha} specifically
considered non-standard energy losses caused by neutrinos and axions
and studied the impact on the TRGB for stellar parameters appropriate
for the globular cluster M5. The color $(V-I)^\TRGB\approx1.8$ is very
similar to the reference value for our empirical
calibrations. Therefore, the modifications caused by particle emission
of  Viaux et al.\ (2013)  carry over to the present case without further modification.

In the degenerate core of a low-mass red giant, the emission of neutrinos would be enhanced if they had a large non-standard magnetic or electric dipole or transition moments. This effect is parameterized by the ``magnetic dipole moment'' $\mu_\nu$ which really sums over all channels, i.e., over all flavors and over magnetic and electric moments, although in practice one expects one contribution to dominate. Moreover, the physical interpretation also depends on the Dirac vs.\ Majorana question that impacts the available final states. In view of the interesting range we finally use the parameter \hbox{$\mu_{12}=\mu_\nu/(10^{-12}\,\mu_{\rm B})$} with $\mu_{\rm B}=e/2m_e$ the Bohr magneton.

After adjusting the zero point according to the critique by S17 discussed in Sec.~\ref{sec:Viaux-Theory}, leading to Eq.~\eqref{eq:Viaux-01}, the updated prediction of V13 in the presence of anomalous neutrino emission is
\begin{equation}\label{eq:mu-prediction}
   M_{I,\mu}^{\rm Theory} = -4.08-\delta M_\mu\pm\sigma_\mu,
\end{equation}
where the anomalous shift and the uncertainty are
\begin{subequations}
\begin{eqnarray}
  \kern-1em \delta M_\mu &=&0.23\Bigl(\sqrt{\mu_{12}^2+0.80^2}-0.80-0.18\mu_{12}^{1.5}\Bigr),
  \\
  \kern-1em \sigma_\mu&=&\sqrt{0.039^2+(0.046+0.0075\mu_{12})^2}.
\end{eqnarray}
\end{subequations}
For $\mu_{12}=0$ the shift $\delta M_\mu$ vanishes whereas the error estimate is the
same as in Eq.~\eqref{eq:Viaux-01}. The modification of the uncertainty with $\mu_{12}$ arises because with increasing brightness, also the color and therefore the bolometric correction changes and also its uncertainty.

\subsection{Bounds on Neutrino Dipole Moments}

The final step in deriving bounds on neutrino dipole moments is to
compare the modified theoretical prediction of
Eq.~\eqref{eq:mu-prediction} with the different empirical
calibrations. Therefore, we consider the offset
\begin{eqnarray}\label{eq:lmc-theory}
  \Delta M_{I}^\TRGB&=&M_I^{\rm Empirical}-M_{I,\mu}^{\rm Theory}
  \nonumber\\
  &=& \Delta M_{I,0}^\TRGB+\delta M_\mu\pm\bar\sigma_\mu\,,
\end{eqnarray}
where $\Delta M_{I,0}^\TRGB$ is the offset at vanishing $\mu$ for the
different calibrations listed in Eq.~\eqref{eq:Viaux-offset}. The
overall uncertainty derives from combining $\sigma_\mu$ and
that of the empirical calibration in quadrature
\begin{equation}\label{eq:sigma-prime}
  \bar\sigma_\mu=\sqrt{\sigma_{\rm Empirical}^2+\sigma_\mu^2}\,.
\end{equation}
Assuming Gaussian errors, these results define a distribution function
\begin{equation}\label{eq:distribution}
  \frac{1}{\sqrt{2\pi}\,\bar\sigma_\mu}
  \exp\left[-\frac{\left(\Delta M_{I}^\TRGB-\Delta M_{I,0}^\TRGB-\delta M_\mu\right)^2}{2\,\bar\sigma_\mu^2}\right]\,,
\end{equation}
which, for fixed $\mu$, can be interpreted as the probability
distribution of possible $\Delta M_I^\TRGB$ values based on the
measurements and theory.

On the other hand, if the only uncertainty derives from $\mu$, the true $\Delta M_I^\TRGB=0$ and the remaining expression gives us the probability distribution for $\mu$. As increased energy losses can only increase the core mass, the $\mu$
distribution encompasses effectively the $\Delta M_I^\TRGB>0$ part of the distribution, so the most restrictive limits arise from those calibrations that favor a dimming of the TRGB
relative to theory. After normalising the integral $\int_{0}^{\infty} d\mu$ of the distribution function to unity, we integrate instead up to those limiting values of $\mu$ that encompass 68.27\% (95.45\%) of the probability and find the limits
\begin{equation}\label{eq:mu-limits}
  \mu_{12}<
  \begin{cases}
    0.75~(1.48) &\quad\hbox{NGC 4258}\\
    0.76~(1.48) &\quad\hbox{LMC (F20)}\\
    0.56~(1.13) &\quad\hbox{LMC (Y19)}\\
    0.58~(1.18) &\quad\hbox{$\omega$ Centauri}
  \end{cases}
\end{equation}
The spread between these different limits is actually quite small for any practical purpose.

\subsection{Axions}

The second case is the emission of axions that are assumed to have a direct Yukawa coupling $g_{ae}$ with electrons. In low-mass red-giant cores, they are primarily emitted by bremsstrahlung in electron-nucleon collisions. Therefore, the radial distribution of axion or neutrino energy losses is different. Standard neutrino cooling pushes the helium ignition point from the center of the core to some non-vanishing radius; explicit examples were shown by Raffelt and Weiss in their Fig.~2 \cite{Raffelt:1992pi}. Non-standard cooling enhances this effect, but for different cooling channels in quantitatively different ways so that the neutrino and axion cases are often treated separately. In view of the interesting range we use the parameter \hbox{$g_{13}=g_{ae}/10^{-13}$}. Another relevant parameter is the axionic fine-structure constant $\alpha_{ae}=g_{ae}^2/4\pi$, often expressed in terms of $\alpha_{26}=\alpha/10^{-26}=g_{13}^2/4\pi$.

Next we repeat the neutrino-dipole exercise for the case of axions for which similar expressions for the brightness shift and its uncertainty apply \cite{Viaux:2013lha}. Actually it turns out that in terms of the parameter $g_{13}$ these results fortuitously are numerically so similar to the dipole ones in terms of $\mu_{12}$ that one could use these parameters almost interchangeably. Specifically we find the new axion limits shown in the last columns of Table~\ref{tab:summary} for the different cases of TRGB calibration.

\section{Discussion and Summary}
\label{sec:conclusion}

In the first part of our paper, we have collected recent extragalactic TRGB calibrations that used geometric distance indicators. First we considered the galaxy NGC4258 at a distance of 7.6~Mpc, whose distance precision through water megamasers has been recently improved by almost a factor of two \cite{Reid:2019tiq}. Then we turned to the Large Magellanic Cloud at $d=50$~kpc, whose TRGB calibrations were performed in F20 \cite{Freedman:2020dne} and Y19 \cite{Yuan:2019npk} in the context of determining the Hubble parameter $H_0$. As shown in Fig.~\ref{fig:summary}, the extragalactic calibrations agree within the quoted uncertainties.

However, there remain unresolved systematic issues concerning the LMC-based calibration because the slight tension between Y19 and F20 mostly derives from different values of the adopted extinction. This discrepancy is uncomfortably large in the context of the Hubble-tension debate, but relatively less important for the new-physics bounds provided here because we need to combine the empirical uncertainties with the currently larger theoretical ones. Still, given the open questions concerning the LMC extinction we use NGC~4258 as the cleanest extragalactic calibration.

Moreover, we have re-calibrated the TRGB using the globular clusters $\omega$~Centauri and M5 using the recent kinematical distance determinations of Baumgardt et al.\ \cite{Baumgardt:2019}, which provide the largest modification besides other minor updates, including extinction. The new calibrations exacerbate the tension between the two globular clusters. The M5 results depend on a star-by-star RGB/AGB discrimination and returning to the brightest stars in M5 we notice that they are variable in brightness and color. Notably the brightest star used by V13 is a semi-regular long-period (108 days)
variable with a large amplitude of variation, so its average properties may have led to its incorrect classification as the brightest RGB star. Therefore, we no longer use M5 for the purpose of TRGB calibration.

The comparison between galactic and  extragalactic calibrations in Fig.~\ref{fig:summary} and Table~\ref{tab:summary} shows fairly good agreement. We note that the uncertainties are also comparable, making both galactic and extra-galactic approaches worth pursuing in the future. We emphasize the complementarity between them: as both of them suffer from different systematic issues, their general agreement provides an extremely useful cross-check on the underlying assumptions and methods adopted during the calibration procedure. We refrain, though, from performing a combination of these calibrations because this task would require an excellent understanding of possible correlations. Moreover, ultimately the main problem are systematics, not statistical fluctuations, so combining heterogenous results would provide artificially small nominal errors.

Improving the empirical TRGB calibration would be interesting in the context of the debate about the Hubble tension. The TRGB-implied $H_0$ value is between the one derived from cosmological data and the local one based on Cepheid distances, being statistically compatible with either one. A refined TRGB calibration might help to clarify this situation. For example, one could use the recent kinematical globular cluster distances to re-calibrate the TRGB using the CMD of many globular clusters, not just the two we have specifically used.

General improvements in the near future are expected with the upcoming Gaia Data Release~3, which will possibly further reduce the uncertainty on the globular cluster distances. While these errors are no longer the dominant source of uncertainty, their further reduction will have a noticeable impact on the final calibrations. Moreover, Gaia DR3 parallaxes will probably allow, for example, to use the halo red giants in the Milky Way for a new TRGB calibration that could play an important role for both the determination of the Hubble parameter $H_0$ and particle bounds.

The latter derive from a comparison of the empirical calibrations with theoretical predictions with or without novel cooling channels. Given the precision of the empirical results, the uncertainties of the stellar models and bolometric corrections have become the dominant sources of uncertainty.
To arrive at our limits, we have closely followed the earlier approach of V13 in that we have combined a large number of systematic issues in quadrature and have interpreted the overall uncertainty as a Gaussian error. Conversely, S17 have preferred to state maximum errors by taking input uncertainties to extremes and adding their effect coherently. Particle bounds based on this approach are roughly comparable to a 2--$2.5\sigma$ error of the former approach, so any substantial conclusion would be similar. Still, the bounds and their formal significance depend on how one deals with systematic effects (``the devil is in the tails'').

A list of adopted uncertainties for the stellar models of V13 is given in their Table~4 and the adopted uncertainty of the bolometric correction in their Eq.~(9), the latter being the largest single source of error in the predicted $M_I^\TRGB$. Another large item is the mixing-length parameter $\alpha_{\rm MLT}$ which is calibrated to reproduce measured properties of the Sun. In particular, it is not adjusted to optimize agreement of the RGB with observations. There are many smaller items concerning microscopic input physics, notably nuclear reaction rates, screening, or conductive opacities. One large item is mass loss on the RGB. The predictions of S17 did not include mass loss at all, whereas V13 included some amount of mass loss (see last line in their Table~4), causing a brightness difference between the predictions on the level of 0.03~mag. On the other hand, the two groups used different BCs and other differences in detail, and in the end the zero points fortuitously agree exactly. This prediction is almost exactly between the two globular-cluster calibrations and within uncertainties agrees with either.

Concerning particle emission, we have reconsidered the red-giant bounds on neutrino dipole moments and the axion-electron coupling strength. The main new ingredients, relative to the earlier studies of Viaux et al.\ \cite{Viaux:2013hca, Viaux:2013lha}, are a correction of the theoretical prediction to account for the treatment of nuclear reaction rates and new TRGB calibrations based on geometrical distance indicators. We propose as reference empirical calibrations for particle bounds those from the galaxy NGC4258 and the globular cluster $\omega$ Centauri. From the former we obtain $\mu_{12}=\mu/10^{-12}\mu_B<0.75\,(1.48)$ at 68\% (95\%) CL, whereas from the latter we get $\mu_{12}<0.58\,(1.18)$.

We have noted that red-giant bounds on neutrino dipole moments $\mu_\nu$ and on the axion-electron coupling $g_{ae}$ in practice are numerically the same in terms of the parameters $\mu_{12}$ and $g_{13}=g_{ae}/10^{-13}$ unless one worries about second-digit precision, so we can summarise previous results in terms of either quantity even if the authors considered only one of them. Previous bounds were reported as $\mu_{12}<2$  \cite{Raffelt:1989xu}, $\mu_{12}<1$ \cite{Castellani:1993hs}, $\mu_{12}<4.5$ (95\%~CL) \cite{Viaux:2013hca}, $g_{13}<4$ (95\% CL) \cite{Straniero:2018fbv}, $\mu_{12}<2.6$  \cite{Arceo-Diaz:2015pva}, and $\mu_{12}<2.2$ \cite{Diaz:2019kim}, to be compared with our most restrictive new bound $\mu_{12}<1.2$ (95\%~CL). Compared with previous results, our bound is the most constraining in terms of specified CL. However, it is also clear that these bounds have not changed very much in around 30~years, reflecting the good agreement between standard stellar evolution theory and observations. What has changed, however, is a more systematic assessment of uncertainties and that the limiting factor now has become stellar evolution theory together with bolometric corrections and no longer the observational data and distance determinations.

A previous analysis by Viaux et al.\ \cite{Viaux:2013hca, Viaux:2013lha}, based on the globular cluster M5, was interpreted as providing a hint for extra cooling \cite{Giannotti:2015kwo}. This weak effect, nominally on the $2\sigma$ level,
can be attributed to an unlucky combination of issues in V13. One is
the dimmer theoretical prediction based on the treatment of screening
in nuclear reaction rates as pointed out by S17. The other is the
identification of the brightest star that turns out to be a
large-amplitude variable so that it may have been misidentified as an RGB star. Red-giant variability near the TRGB exacerbates the difficulty of AGB/RGB separation.

Recent geometric distance determinations have vastly improved the empirical TRGB calibrations. Further progress, driven by the Hubble-tension debate, may well derive from a better systematic understanding of LMC extinction, from using many galactic globular clusters with good geometric distances, and from Gaia DR3 parallaxes of galactic halo red giants. For the purpose of particle bounds, the limiting factors are becoming theoretical models together with mapping between their predictions and observational data in the form of the bolometric corrections. It would be important to develop a better understanding of the differences between different prescriptions and their uncertainties.

\section*{Acknowledgements}

We thank Achim Weiss for discussions on the theoretical calibration of Serenelli et al.\ \cite{Serenelli:2017} and for pointing us to the literature on extragalactic TRGB calibrations. We also thank Gonzalo Alonso-\'Alvarez and Marta Reina-Campos for discussions and for pointing us to the kinematical distance determinations for galactic globular clusters \cite{Baumgardt:2019}. Particular thanks go to the anonymous referee for many   constructive suggestions that have strongly improved our paper. We acknowledge partial support by the Deutsche For\-schungs\-ge\-mein\-schaft (DFG) through Grants No.\ EXC 2094–390783311 (Excellence Cluster ``Origins'') and SFB 1258 (Collaborative Research Center ``Neutrinos, Dark Matter, Messengers''). This work was initiated by discussions at the MIAPP workshop ``Axion Cosmology'' (17 February--13 March 2020, Garching, Germany), see \url{https://www.munich-iapp.de/activities/activities-2020/2020-1}.

\bibliographystyle{bibi}
\bibliography{Bibliography}

\end{document}